# A Managed Tokens Service for Securely Keeping and Distributing Grid Tokens

*Shreyas* Bhat[1,*], *Dave* Dykstra[1]

[1] Computational Science and Artificial Intelligence Directorate, Fermi National Accelerator Laboratory, Batavia, IL, USA

**Abstract.** Fermilab is transitioning authentication and authorization for grid operations to using bearer tokens based on the WLCG Common JWT (JSON Web Token) Profile. One of the functionalities that Fermilab experimenters rely on is the ability to automate batch job submission, which in turn depends on the ability to securely refresh and distribute the necessary credentials to experiment job submit points. Thus, with the transition to using tokens for grid operations, we needed to create a service that would obtain, refresh, and distribute tokens for experimenters' use. This service would avoid the need for experimenters to be experts in obtaining their own tokens and would better protect the most sensitive long-lived credentials. Further, the service needed to be widely scalable, as Fermilab hosts many experiments, each of which would need their own credentials. To address these issues, we created and deployed a **Managed Tokens service**. The service is written in Go, taking advantage of that language's native concurrency primitives to easily be able to scale operations as we onboard experiments. The service uses as its first credentials a set of kerberos keytabs, stored on the same secure machine that the Managed Tokens service runs on. These kerberos credentials allow the service to use **htgettoken** via **condor_vault_storer** to store vault tokens in the **HTCondor** credential managers (credds) that run on the batch system scheduler machines (HTCondor schedds); as well as downloading a local, shorter-lived copy of the vault token. The kerberos credentials are then also used to distribute copies of the locally-stored vault tokens to experiment submit points. When experimenters schedule jobs to be submitted, these distributed vault tokens are used to access a **HashiCorp Vault** instance (run separately from the Managed Tokens service), and previously stored refresh tokens there are used to obtain the bearer token that is submitted with the job. We will discuss here the design of the Managed Tokens service, including elaborating on certain choices we made with regards to concurrent operations, configuration, monitoring, and deployment.

## 1 Background

For the past few years, Fermilab has been transitioning authentication and authorization for grid operations away from using X.509 user certificates (with additional embedded attributes from a Virtual Organization Management System (VOMS) server [1]) to using bearer tokens (a.k.a. access tokens) based on the WLCG Common JWT (JSON Web Token) Profile [2]. Many Fermilab experimenters rely on the ability to run complex workflows in batch jobs, which requires the ability to automate job submission. To do so, experiments previously utilized X.509 Proxy Certificates that were automatically generated and renewed. That functionality was provided by a service run by Fermilab called the **Managed Proxies service**.

---

[*] Corresponding author: sbhat@fnal.gov

With the transition to bearer tokens for job submission, we needed to replicate that service to provide experiments with bearer tokens at their job submit points. It would have been possible for individual experiments to create their own scripts to obtain their own bearer tokens, but due to the sheer number of experiments run at Fermilab, each with their own unique credentials, we decided that it would be more prudent and serve the experiments better to create a service to handle the generation and distribution of these tokens on their behalf. This would also enable us to implement the best security practices for handling various long-lived secrets required for this process, as well as the distribution of the tokens.

Thus, by developing, deploying, and operating this **Managed Tokens service** for the distribution of bearer tokens, we intended to provide Fermilab's experiments with a reliable, sophisticated, performant service that would abstract away the credential-obtaining process, and eliminate the duplicated effort that would otherwise certainly be required.

## 2 Architecture

### 2.1 Requirements

**HTCondor** [3] is the batch system that is used by Fermilab experiments for High-Throughput Computing tasks. In the HTCondor pool, the responsibility of distributing and renewing tokens inside the user jobs lies with the scheduler daemon, or **schedd**. On an HTCondor scheduler machine along with a schedd, a credential daemon (**credd**) is also installed. It is the credd that handles storing credentials locally and refreshing them as needed for running jobs.

In a standard OAuth 2.0 workflow [4], a user obtains a short-lived **bearer token**, which authorizes them to perform the desired task, and a longer-lived **refresh token**, which gives the user the ability to refresh the bearer token. At Fermilab, rather than task the users with keeping track of the refresh token, we decided to use an existing open-source secrets management package, **HashiCorp Vault** [5], to store the refresh tokens. We instead provide users with a different token, a **vault token**, which authorizes them to access the refresh token in Vault whenever the bearer token needs to be refreshed. Further, the HTCondor credd also supports using Vault to keep bearer tokens refreshed via the **condor-credmon-vault** package [6]. Details of the condor-credmon-vault package, how it integrates with HTCondor, and the general Vault-HTCondor architecture can be read in [6] and will not be discussed further here.

The user, then, has access to two tokens – a bearer token, which in the Fermilab implementation, is valid for 3 hours, and a vault token, which is valid for 7 days. We decided that rather than distribute bearer tokens, it would be more user-friendly and fault-tolerant to distribute vault tokens to the experiments. Users can then use a tool like **htgettoken** [6] to obtain a bearer token. As a part of the wider transition to using JWTs, many of Fermilab's grid utilities, including **jobsub** [7] for grid job submission and **ifdh** [8] for file transfers, were updated to automatically run htgettoken to obtain a bearer token for the user using an existing vault token or kerberos credential. Most users use one of these latter tools, rather than running htgettoken directly.

In addition, we wanted to ensure that the service was well-monitored so that, in addition to making sure the service ran smoothly, we could continually improve its reliability and performance. Thus, we needed to incorporate all of the elements of the modern observability stack – logs, metrics, and tracing – into this service.

With all of these requirements in mind, we set out to build a robust service that could use the existing infrastructure to obtain vault tokens, store them in the HTCondor credds, distribute them to experiment submit points, and ensure that those vault tokens were always valid for experiments to use to run grid operations.

## 2.2 Selection of Implementation Language

In implementing this service, we wanted to select a language that was easy to develop in and supported easy scalability. Since Fermilab hosts many experiments, each with possibly multiple different permission sets (capability sets), and each experiment can submit jobs to multiple HTCondor schedds, scalability became a top priority. Due to its ability to easily launch and synchronize multiple concurrent operations, we chose **Go** [9] as the language to implement Managed Tokens. We were able to quickly design and build a system that spins out a worker thread for each experiment and operation, and synchronize their actions from the main executable, **token-push**.

## 2.3 System Architecture

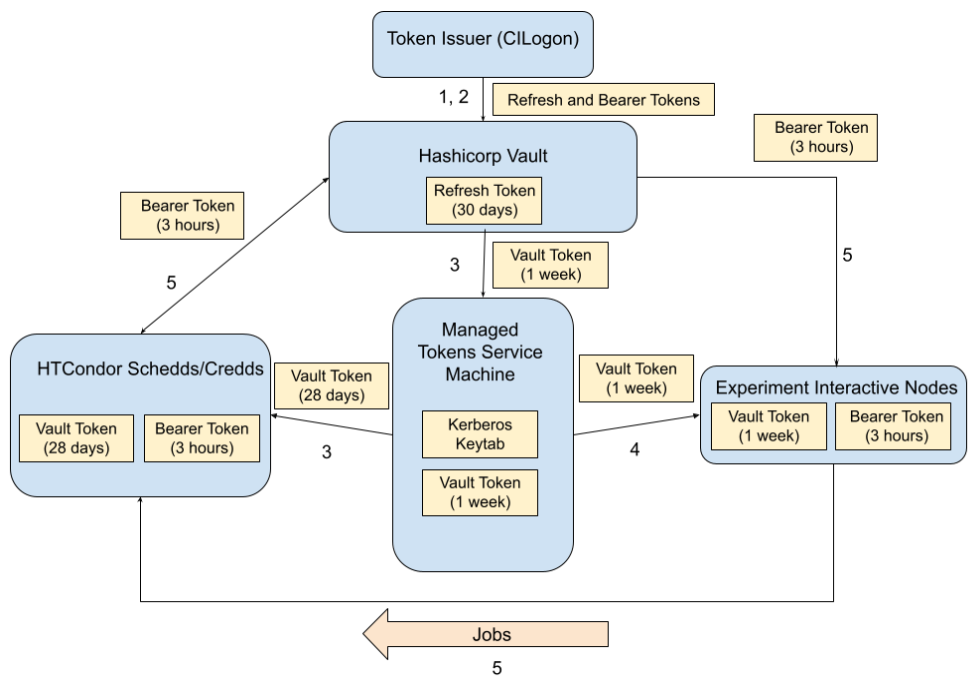

**Fig. 1.** Architecture of Managed Token service within Grid Infrastructure

Figure 1 shows how the Managed Tokens service server interacts with the various components of the grid infrastructure. The steps are as follows:
1. For the initial onboarding, the operator must generate the refresh token and store it in Vault using OIDC authentication (see section 2.3.1).

2. Otherwise, the Managed Tokens server uses a Kerberos credential and a vault token to contact Vault and possibly renew the vault and refresh tokens.
3. This operation results in a vault token being stored in the HTCondor credd and a shorter-lived vault token being stored on the Managed Tokens server.
4. The Managed Tokens service distributes copies of the shorter-lived vault tokens to the various experiment submit points (interactive nodes in Fig. 1).
5. Experiments can use these vault tokens to generate bearer tokens to run grid jobs, transfer files, or for other token-enabled grid operations. When grid jobs are submitted, the HTCondor schedds will use the vault tokens from (3) to obtain bearer tokens for the jobs to use.

The Managed Tokens service runs steps 1-4 above for each registered capability set. In the Fermilab grid infrastructure, capability sets map to special UNIX users. However, we did not want to add every user account to the Managed Tokens server. Thus, we designed the service to run from a single service account that has access to the various Kerberos credentials.

The service itself is split into two executables, **refresh-uids-from-ferry** and **token-push**. The former simply periodically queries the Fermilab instance of **FERRY** [10] and keeps an on-disk database of the UID for each UNIX account/capability set. The latter, token-push, is the main executable of the Managed Tokens service. Its high-level program flow is shown in Figure 2.

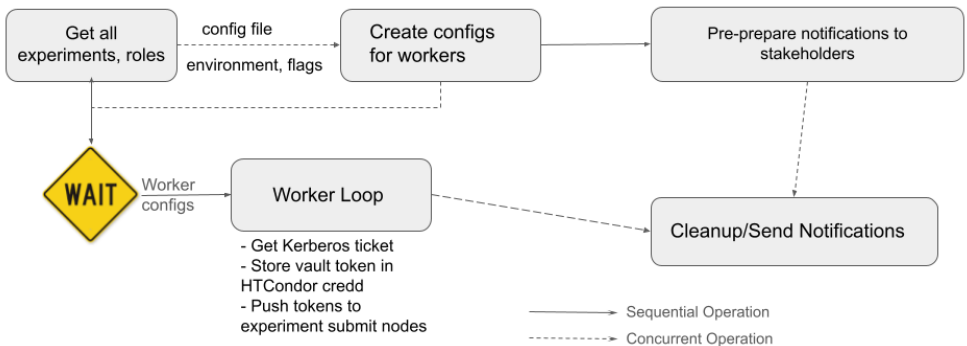

**Fig. 2.** General program flow for token-push, the main Managed Tokens service executable

As can be seen in Figure 2, the core of the general design of the token-push program is a thread pool pattern (see "Worker Loop" in Fig. 2), where the main thread, (called a goroutine in Go parlance) spins out and orchestrates worker goroutines to accomplish the various tasks and handles notifications, which are aggregated in a separate goroutine. We will now turn our attention to each task that is handled by the workers.

There are alternative designs possible that would be simpler, for example keeping the refresh token itself on a managed token server but, given that we already had the Vault server in the architecture set up to use Kerberos, this was the simplest extension.

### 2.3.1 Kerberos Tickets and Storage of Vault Tokens

The initial authentication to generate a refresh token and store it in Vault is done by an operator of the Managed Tokens service using OIDC authentication and a web browser. The resultant refresh token is stored in a path in Vault that matches the Kerberos credential used for renewal of Vault access. This Kerberos credential is a specially-generated "robot" credential that may be used indefinitely and thus is stored securely on the Managed Tokens service server. As explained in [6], the name of this credential is usually of the form "user/type/machinename", where "user" is the username of the intended experiment user, "type" is an arbitrary string denoting the purpose of the credential, and "machinename" is the fully qualified hostname of the machine using the credential – in this case the Managed Tokens server. While the Managed Tokens service assumes a convention for all three of these fields in the Kerberos credential name, the name is configurable.

The Managed Tokens service uses **condor_vault_storer** [6] from the condor-credmon-vault package to communicate with Vault, and to store and obtain vault tokens. When a new experiment onboards to the Managed Tokens service, no refresh token exists in Vault, and so condor_vault_storer will prompt the operator to complete the OIDC flow discussed in [6]. This means that the operator must be registered with the token provider as having access to the capabilities the final bearer token will need. This initial authorization will store a 28-day vault token in the various HTCondor credds and a 7-day vault token on the Managed Tokens server. After the initial onboarding, the Managed Tokens service runs condor_vault_storer periodically, and Kerberos credential and vault token are sufficient for condor_vault_storer to refresh the vault and refresh tokens as needed, indefinitely.

condor_vault_storer was designed to be run by different users on the same machine, for a single schedd; not by a single service account concurrently managing multiple vault tokens for multiple schedds. There are two effects of that limitation: 1) condor_vault_storer cannot be run concurrently, thus the running of condor_vault_storer is the main performance bottleneck; and 2) Each time condor_vault_storer refreshes the on-disk vault token, that token must be moved to a different location on disk, to avoid collision with other experiments' vault tokens.

### 2.3.2 Distribution of Vault Tokens to Final Destinations

Once the vault tokens have been obtained, the token-push executable copies the token files to the experiment submit points (interactive nodes). This is done by running concurrent **rsync** [11] commands within goroutines to distribute two copies of each vault token to the destination machine's /tmp directory: one to the location where HTCondor expects the vault token to be, and one to the location where htgettoken expects it to be. The reason these two locations are different is because HTCondor, through condor_vault_storer, tries to separate vault tokens by UID, issuer, and role, whereas htgettoken only distinguishes vault tokens by user, following the WLCG bearer token discovery specification [12].

### 2.3.3 Notifications

While the preceding actions are taking place, token-push will concurrently start a set of listener threads that aggregate any errors into notifications that get sent to experiment stakeholders. Due to the various transient issues that could be affecting the numerous experiment submit points across the Fermilab Computing Facility, we settled on notifying experiment stakeholders when three consecutive attempts to distribute a particular vault token to a particular submit point fail. This frequency of alerting seems to strike a reasonable balance between providing good advance notice to the Managed Tokens service operators and experiment stakeholders that there may be an issue with either the service side or the experiment submit point; and not overwhelming either with "alarm fatigue".

## 3 Deployment

The Managed Tokens service is deployed via an RPM which installs the pre-compiled Go executables along with a stock crontab that runs the executables, and a YAML configuration file. At this time, a single instance of the service is sufficient to distribute vault tokens to the Fermilab experiments. It is deployed on a secure server, which also stores the initial Kerberos credentials mentioned previously. To ensure reproducible deployments, we are currently in the process of automating the deployment steps to run under **Puppet** [13].

## 4 Observability

Given that the Managed Tokens service sits between many other services and components, being able to monitor the service at varying levels of detail is extremely important. To that end, this service implements the standard observability layers of logging, metrics, traces, and dashboards:

• The service logs are stored on disk, and all INFO and higher-level events are also sent to a **Grafana Loki** [14] instance that runs on Fermilab's central monitoring infrastructure, **Landscape** [15]. This proved extremely useful in one case recently, when we were able to use the Loki logs to quickly pinpoint an issue to an experiment submit point environment by comparing the logs from different experiment vault token pushes.

• **Prometheus** [16] Metrics are collected regarding operation successes and failures along with other performance measures. Since the Managed Tokens service runs as a batch job rather than a continuously running service, the metrics are pushed to a **Prometheus Pushgateway** [17]**,** and this gateway is scraped by the Landscape Prometheus server.

• **OpenTelemetry**-compliant [18] traces are collected and are viewable through a **Jaeger** [19] instance. In troubleshooting we frequently use these as a quick way to drill down to exactly which component of the service is causing issues before consulting the logs for further detail.

• We built **Grafana** [20] dashboards, again hosted on Landscape, that allow operators and stakeholders to ascertain the performance of the service. Figure 3 shows the main operator dashboard for the Managed Tokens service, built from Prometheus metrics. Many of these dashboards also will send alerts to operators if there is an issue with the service that requires intervention.

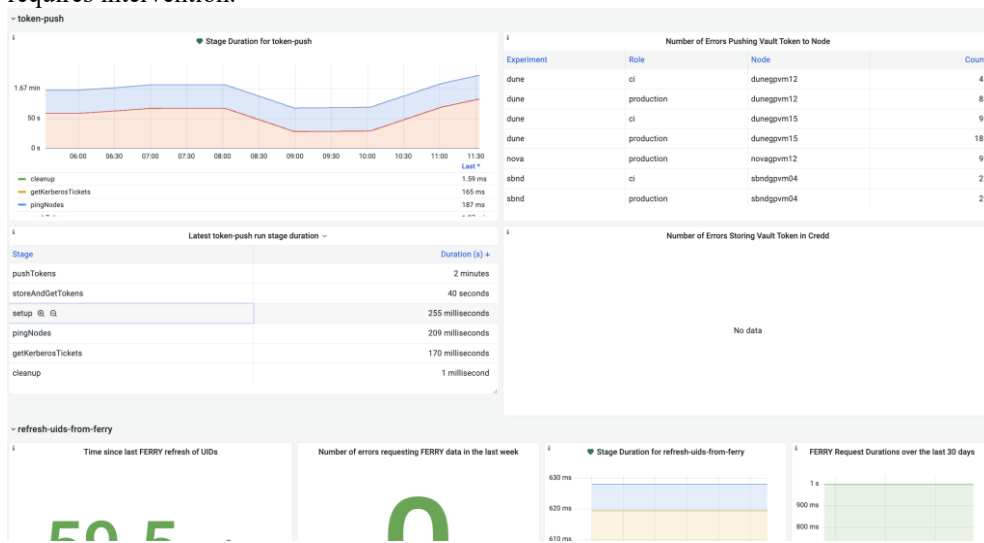

**Fig. 3.** Managed Tokens Service Health Dashboard for operators in Grafana

## 5 Lessons Learned and Conclusions

The Managed Tokens service has been running in production since November 2022, with very few major issues. It has software dependencies that were rapidly evolving at the same time this product was being developed. Taking the time to design the architecture of the product before writing any code, rather than only ensuring it met stakeholder requirements, allowed us to quickly adapt to this changing landscape of the token infrastructure at Fermilab (this is expanded upon in another paper in these proceedings [21]), whether through bugfixes, changes, or added features.

Since Managed Tokens was released with a strong monitoring infrastructure, it allowed us to scale with the confidence that we would be able to identify any system-wide issues before they disrupted service to stakeholders. We continue to make small changes to the code base to ensure better service and reliability.

Currently, the service is designed to be usable by anyone hosting experiments that use htgettoken to obtain access tokens, Kerberos-authenticated HashiCorp Vault to store refresh tokens, and HTCondor as their batch system. However, we do have plans in the future to broaden the libraries to allow for wider support for different infrastructures.

The Managed Tokens software is open-source, available at https://github.com/fermitools/managed-tokens. We welcome any contributions, pull requests, and suggestions.

## 6 Acknowledgement